\documentclass[smus]{snow2ex}
\usepackage{psfig}
\usepackage{equations}
\usepackage{latexsym}
\let\Diamond=\diamond
\let\square=\Box
\def\nicefrac#1#2{{\textstyle{#1\over#2}}}
\newcommand {\mz} {$M_{Z}$}
\newcommand {\mzsq} {$M_Z^2$}
\newcommand {\et} {$E_T$}
\newcommand {\qsq} {$Q^2$}
\newcommand {\xf} {$x$}
\newcommand {\epem}   {$e^+e^-$}
\newcommand {\ppr}      {$pp$}
\newcommand {\pp}       {$p\overline{p}$}
\newcommand {\alphas} {$\alpha_{S}$}
\let\als=\alphas
\newcommand {\alphasq} {$\alpha_{S}$(\qsq)}

\newcommand {\ftwo} {$F_{2}$}
\newcommand {\ftwoxq} {$F_{2}$(\xf,\qsq)}

\newcommand {\fthreexq} {$F_{3}$(\xf,\qsq)}
\newcommand {\fthree} {$F_{3}$}
\newcommand {\xfthree} {{\xf}$F_{3}$}
\newcommand {\nubar} {$\overline{\nu}$}

\newcommand {\mwsq}  {$M_{W}$}
\newcommand {\gone} {$g_1$}
\newcommand {\gonexq} {$g_1$(\xf,\qsq)}

\newcommand {\qq} {$q\overline{q}$}
\newcommand {\toptop} {$t\overline{t}$}
\newcommand {\gdist} {$g$(\xf,\qsq)}
 
\newcommand{\cc}{c\bar{c}}
\newcommand{\bb}{b\bar{b}}

\newcommand{\al}{\alpha}
\newcommand{\GeV}{{\rm GeV}}

\newcommand{\ms}{\overline{{\rm MS}}}
\newcommand{\Tr}{{\rm Tr}}

\newcommand{\be}{\begin{equation}}
\newcommand{\ee}{\end{equation}}
\newcommand{\gtap}{{\raise.3ex\hbox{$>$\kern-.75em\lower1ex\hbox{$\sim$}}}}
\newcommand{\ltap}{{\raise.3ex\hbox{$<$\kern-.75em\lower1ex\hbox{$\sim$}}}}

\newcommand{\amsmz}{$\alpha_{\overline{{\rm MS}}}(M_{Z})$}
\newcommand{\afmsmz}{$\alpha^{(5)}_{\overline{{\rm MS}}}(M_{Z})$}

\newcommand{\PRD}[3]{{\em Phys. Rev.} {\bf D{#1}} (19{#2}) {#3}}

\newcommand{\PRL}[3]{{\em Phys. Rev. Lett.} {\bf {#1}} (19{#2}) {#3}}
\newcommand{\PLB}[3]{{\em Phys. Lett.} {\bf B{#1}} (19{#2}) {#3}}
\newcommand{\NPB}[3]{{\em Nucl. Phys.} {\bf B{#1}} (19{#2}) {#3}}
\newcommand{\NPBproc}[3]{{\em Nucl. Phys.} {\bf B} (Proc. Suppl.)
           {\bf {#1}} (19{#2}) {#3}}
\newcommand{\ARNPS}[3]{{\em Annu. Rev. Nucl. Part. Sci.}
           {\bf #1} (19{#2}) {#3}}

\begin{document}
 
\setcounter{footnote}{1}
 \def\thefootnote{\fnsymbol{footnote}}
 \def\@makefnmark{\hbox
    to 5pt{$^{\@thefnmark}$\hss}}
       \large
\thispagestyle{empty}
\hsize=\textwidth
\rightline{SLAC--PUB--7371}
\rightline{ILL--TH--96--10}
\rightline{SCIPP 96/56}
\rightline{UCRHEP--E180}
\rightline{hep-ex/9612012}
\rightline{December 1996}
\vspace*{2cm}
\begin{center}
{\LARGE\bf Prospects for the Precision Measurement
of \boldmath$\alpha_s$} 
\\

\vskip2pc

   P. N. Burrows\\ {\it Massachussetts Institute of Technology} \\[2pt]
   L. Dixon \\{\it Stanford Linear Accelerator Center} \\[2pt]
   A. X. El-Khadra\footnote{Subgroup conveners}\\[2pt]
            {\it University of Illinois, Urbana} \\[2pt]
   J. W. Gary$^\dag$\\ {\it University of California, Riverside}   \\[2pt]
   W. Giele\\ {\it Fermi National Accelerator Laboratory} \\
   D. A. Harris\\ {\it University of Rochester}              \\[2pt]
   S. Ritz\\ {\it Columbia University}                  \\[2pt]
   B. A. Schumm$^\dag$\\ {\it University of California, Santa Cruz}\\
 
\vspace*{1.25cm}
\begin{minipage}{15cm}
 \centerline{ABSTRACT}
\vskip1pc
The prospects for the measurement of the strong coupling constant
{\amsmz} to a relative uncertainty of $1\,\%$ are discussed.
Particular emphasis is placed on the implications
relating to future High Energy Physics facilities.
\end{minipage}
 
\vspace*{1.5cm}
To be published in      \\
Proceedings of the 1996 DPF/DPB Summer Study on \\
New Directions for High-Energy Physics (Snowmass 96) \\
\vfill
This work has been supported in part by the Sloan Foundation and
U.S. Department of Energy grants
DE-FG03-94ER40837,
DE-AC03-76SF00515,
DE-FG02-91ER40677,
DE-AC02-76ER03069,
DE-FG02-91ER40685,
DE-FG03-92ER40689.
 \end{center}

       \clearpage
\normalsize
\setcounter{page}{1}
\title{Prospects for the Precision Measurement
of \boldmath$\alpha_s$}
 
\author{
   P. N. Burrows\\ {\it Massachussetts Institute of Technology} \\[2pt]
   A. X. El-Khadra\thanks{Subgroup conveners}\\[2pt]
            {\it University of Illinois, Urbana} \\[2pt]
   W. Giele\\ {\it Fermi National Accelerator Laboratory} \\
   S. Ritz\\ {\it Columbia University}                  \\[2pt]
\and
   L. Dixon \\{\it Stanford Linear Accelerator Center} \\[2pt]
   J. W. Gary$^\dag$\\ {\it University of California, Riverside}   \\[2pt]
   D. A. Harris\\ {\it University of Rochester}              \\[2pt]
   B. A. Schumm$^\dag$\\ {\it University of California, Santa Cruz}\\
}
\maketitle
 
\thispagestyle{empty}\pagestyle{plain}

\begin{abstract}
The prospects for the measurement of the strong coupling constant
{\amsmz} to a relative uncertainty of $1\,\%$ are discussed.
Particular emphasis is placed on the implications
relating to future High Energy Physics facilities.
\end{abstract}
 
\section{Introduction}

Quantum Chromodynamics (QCD), the theory of the
strong interaction, has a single free parameter, the strong
coupling  {\alphas}.
The coupling depends on the renormalization scheme and the energy
scale, $Q$.
Once {\alphas}($Q$) is determined from an experimentally measured process,
any other process mediated by the strong interaction can be calculated to
arbitrary accuracy, at least in principle.
Most determinations of {\alphas} are based on perturbative QCD,
where it is conventional to evaluate the coupling in the $\ms$ scheme,
which is only defined in perturbation theory.
Furthermore, it is also customary to choose the $Z^0$ mass, {\mz},
as the reference scale.
We shall adhere to these conventions and quote, for the most part,
{\amsmz} in our discussions.
 
A precise measurement of {\alphas} is motivated
by a number of considerations:
 
\noindent 1.\
The couplings of the electroweak theory, $\alpha_{\rm em}$ and
$\sin^2\theta_W$, have been determined with a precision
of about $0.1\,\%$.
In contrast, the strong coupling is presently known only to
about $5\,\%$.
It is pertinent to improve the accuracy with which
the strong coupling has been measured
in order to place it on a more equal basis with
respect to the other interactions.
For example, the current accuracy of {\alphas} measurements
is one of the main limitations on Standard Model
electroweak tests at LEP and SLC~\cite{bib-lepewwg96}.
In addition, attempts to constrain Grand Unified models,
from the convergence of the standard model SU(3), SU(2) and U(1)
couplings at a Grand Unification Scale, are similarly limited by
the accuracy with which {\alphas} has been measured.
 
\noindent 2.\
QCD with its one parameter, {\alphas}, must account for
the rich phenomenology which is attributed to the strong interaction,
including perturbative and nonperturbative phenomena.
A fundamental test then of QCD is the determination of {\alphas} from
experimental measurements which probe complementary processes.
This test is only meaningful if the values of {\alphas}
being compared have been measured with similar, good accuracy.
 
\noindent 3.\
The QCD $\beta$-function (which is known to three loops in
the $\ms$ scheme) determines the evolution of the coupling.
Accurate measurements of {\alphas} over a wide range of
momenta provide an additional fundamental test of the theory.
Tests of the QCD $\beta$-function constrain physics beyond the standard
model, in particular models with additional colored particles.
Measurements of the energy dependence of
observables in a single experiment,
such as jet variables at {\epem} or {\pp} colliders,
can also test the QCD $\beta$-function.
 
The last two reasons given above for an accurate measurement
of {\alphas} emphasize that it is not sufficient to determine
{\alphas} using a single method,
but that precise measurements are necessary using
different processes and widely different {\qsq} values.
 
For the presentation here, we consider
the prospect to measure {\amsmz} with $1\,\%$ accuracy.
We attempt to identify those methods which offer the
greatest potential for such precision.
 
Figure 1~\cite{avalp} presents a summary of the
most accurate measurements of {\alphas} which
are currently available.
Measurements performed at {\qsq} scales different
from {\mzsq} have been evolved to {\qsq}={\mzsq}
using the three loop QCD $\beta$-function (in the $\ms$ scheme).
All determinations of {\alphas} receive contributions from
theoretical systematic errors. These are, in many cases, the dominant
sources of uncertainty. In general, they are difficult to
estimate. In determinations based on perturbative QCD, sources of
such errors are the truncation of the perturbative series and
nonperturbative effects (such as hadronization).
Most of the perturbative calculations have been carried out to
next-to-leading order (NLO), and, in a few cases, to
next-to-next-to-leading order (NNLO).
The sources of theoretical uncertainty in determinations based on lattice
QCD are discussed in section~III.
In a few cases, {\alphas} results are limited by experimental uncertainties.
 
\begin{figure}[htbp]
\centering
\centerline{\psfig{file=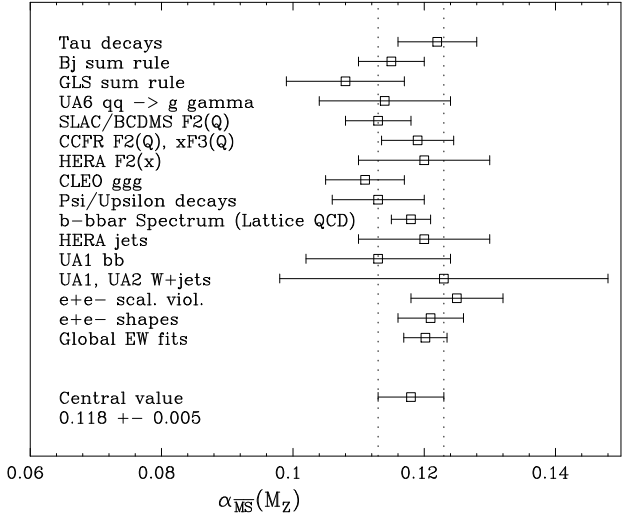,width=8.5cm,angle=90}}
\caption{Summary of current accurate measurements of \als, by technique.
The \als\ measurements are based on perturbative QCD, except where otherwise
noted.
}
\label{fig:one}
\end{figure}

Theoretical uncertainties are in general not 
gaussian-distributed and are estimated from
a variety of different methods.
Consequently, the correlations between
different {\alphas} measurements are difficult to estimate.
Given this difficulty, there is not a unique procedure to
define a world average for the results shown in
Figure 1.
A number of proposals for
world averages exist~\cite{pdg,avalp,bethke}. We shall use~\cite{avalp}
\be
    \al_{\ms}({M}_{Z}) = 0.118 \pm 0.005
\ee
as our nominal value for the world average.
 
In conducting this study, we considered a wide range
of approaches to the measurement of {\amsmz},
eventually identifying four methods which exhibit
the potential to yield results with about $1\,\%$ precision.
The existence of perturbative calculations to at least
next-to-next-to-leading order is a prerequisite for $1\,\%$ accuracy.
A number of such calculations are in progress~\cite{nnlo}, 
and we shall assume that they will be available for the
experimental measurements in question.
 
The four methods are:
(1)~the {\qsq} evolution
of the parity violating structure function {\xfthree},
(2)~the Gross-Llewellyn-Smith sum rule,
(3)~spin averaged splittings in the $\Upsilon$ and $\psi$ systems,
and (4)~hadronic observables in {\epem} annihilations.
Items (1) and (2) in the above list are measured in
deep inelastic neutrino scattering experiments.
Item~(3) is based on lattice QCD, all other methods use
perturbative QCD.
In addition to these four approaches, we found that two
other methods, the {\qsq} evolution of the parity
non-violating structure function {\ftwo} at high~{\xf},
and the jet {\et} spectrum in high energy
proton-(anti)proton collisions,
offer the possibility to determine {\alphas}
with good accuracy in regions of {\qsq} which are
complementary to those of the other measurements.
The ultimate accuracy with which {\alphas}
can be determined using these last
two techniques is uncertain at present, however.
Therefore we do not include them in our final list
of techniques which might yield an {\alphas} result
with $1\,\%$ precision.
 
The remainder of this report is devoted to a
presentation of the various methods we considered
for an {\alphas} measurement,
with a particular emphasis on the implications
for future High Energy Physics facilities.

\section{Deep Inelastic Scattering}
 
Measurements of nucleon structure functions
from the deep inelastic scattering (DIS)
of a lepton on a nuclear target have yielded
some of the most precise results for
the strong coupling constant~{\alphas}.
The field of nucleon structure functions remains
very active,
with major experiments in operation
at CERN, DESY, Fermilab and SLAC.
New structure function data,
extending the measurements to previously unmeasured
regions of the kinematic variables {\xf} and {\qsq}
and utilizing polarized targets and probes,
have recently become available.
These programs are expected to continue for
at least the rest of this decade.
 
In this section,
we assess which of these new data
have the potential
to yield an {\alphas} measurement with~$1\,\%$ accuracy.
We do not review the formalism of structure functions
or provide more than an indication of the
methods used to determine {\alphas} from them.
References to existing literature with
such information are given below where deemed appropriate.
 
\subsection{DIS Nucleon Structure Functions}
\label{sec-dis-sfs}
 
The basic kinematic variables of DIS are
the {\qsq} of the interaction,
given by the difference in 4-momentum squared
between the outgoing and incoming leptons,
and the Feynman variable {\xf} defined by
{\xf}={\qsq}$/(2M(E$-$E^{\prime}$)),
where $M$ is the mass of the target nucleon,
with $E$ and $E^{\prime}$ the energies
of the initial- and final-state leptons, respectively,
as measured in the laboratory frame.
In the quark--parton model,
{\xf} is interpreted to be the fraction of the
nucleon's energy carried by the struck parton.
Experiments in DIS measure the energy and scattering
angle of the final-state lepton and/or
recoiling hadronic system.
The lepton probes are either electrically charged
(electron $e$ and muon $\mu$ probes)
or neutral
(neutrino $\nu$ or antineutrino~{\nubar} probes).
The dominant mechanism for charged
lepton scattering is single photon exchange
in the $t$-channel between the lepton and nucleon system,
while that for $\nu$ or {\nubar} scattering is
single $W^{\pm}$ exchange.
For {\qsq} values which approach or exceed {\mwsq},
$W^{\pm}$ and $Z^0$ exchange become important
for charged lepton scattering.
Nonperturbative QCD corrections to single-parton scattering
contribute higher twist terms to the cross sections,
which scale like (1/{\qsq})$^{\mathrm{n}}$ (n=1,2,3$\cdots$)
and are important at low {\qsq}
(typically \mbox{{\qsq}$<$4-5~GeV$^2$).}
 
Of the many structure functions necessary to
describe DIS cross sections in their most general form,
only three are candidates for a precise
measurement of~{\alphas}:
the structure functions
{\ftwoxq}, {\fthreexq} and {\gonexq}.
{\ftwo} is measured from the neutral current
cross section for unpolarized charged leptons
to scatter from unpolarized targets
and from the sum of the charged current cross sections
for neutrinos $\nu$ and antineutrinos {\nubar}
to scatter from unpolarized targets.
{\fthree} is measured from the difference between
the charged current $\nu$ and {\nubar} cross sections
for scattering from unpolarized targets.
{\gone} is measured from the asymmetry in the
cross sections for longitudinally polarized charged
leptons to scatter from polarized targets if the
beam and target polarizations are parallel,
compared to the case that they are antiparallel,
and from the corresponding asymmetry for targets which
are polarized perpendicular to the beam
polarization directions~\cite{bib-g1meas}.
The techniques that have been used to determine {\alphas}
using {\ftwo}, {\fthree} and {\gone} are
mentioned in the following section.
 
Structure functions can be resolved into color singlet
and color non-singlet components.
In QCD,
the singlet and non-singlet terms evolve
differently with {\qsq}~\cite{bib-gribov}.
The singlet component receives a contribution from
gluon splitting into {\qq} pairs.
As a consequence,
the {\qsq} evolution of the color singlet
term depends not only on the running coupling constant
{\alphasq} but also
on the probability for gluon splitting,
given by the gluon distribution function {\gdist}.
This dependence on {\gdist} is not important
if {\xf} is larger than about 0.25 because the
probability for gluon splitting at large {\xf} is small.
Gluon splitting does not contribute to
the non-singlet component of the structure functions:
the {\qsq} evolution of this term
depends on {\alphasq} only,
irrespective of the {\xf} range.
Depending on the nature of the target
(e.g. deuterium d$_2$ or hydrogen h$_2$),
{\ftwo} is either a pure singlet or a mixture
of a singlet and a non-singlet,
whereas {\fthree} is always a pure non-singlet.
 
\subsection{Methods used to Determine {\alphas}
from DIS Structure Functions}
\label{sec-dis-methods}
 
The following methods have been used
to determine {\alphas} using the
{\ftwo}, {\fthree} and {\gone} structure functions:
\begin{enumerate}
  \item the {\qsq} evolution of {\ftwo} at high {\xf} values,
        measured in charged lepton
        scattering ($4\,\%$)~\cite{bib-virchaux1};
  \item the same method as given in item 1,
        except at low {\xf} values ($11\,\%$)~\cite{bib-nmc93};
  \item the {\qsq} evolution of {\fthree} multiplied
        by the kinematic variable~{\xf},
        i.e. the evolution
        of {\xfthree}~($5\,\%$)~\cite{bib-ccfr93};
  \item the Gross, Llewellyn-Smith (GLS) sum rule,
      based on {\fthree} at fixed {\qsq},
      integrated over all {\xf} values
      ($7\,\%$)~\cite{bib-ccrf93b};
  \item the Bjorken sum rule,
      based on the difference between the
      {\gone} structure functions of protons and neutrons
      at fixed {\qsq},
      integrated over all {\xf} values
      ($4\,\%$)~\cite{bib-ellis95};
  \item the shape of {\ftwo} from charged lepton scattering
      in the limit of very large {\qsq} and
      very small {\xf} values ($9\,\%$)~\cite{bib-ball95}.
\end{enumerate}
The reference given after each item refers to the most
precise result available for the method.
This precision itself,
$\Delta${\amsmz}/{\amsmz},
is given by the number in parentheses,
where {\amsmz} is the value of {\alphas} after it
has been evolved to the scale of the $Z^0$ mass
and $\Delta${\amsmz} is the corresponding uncertainty
including statistical and systematic terms.
 
In addition to the methods listed above,
DIS experiments have measured {\alphas}
using one technique which is not based on
structure functions:
the measurement of {\alphas}
using jet rates~\cite{bib-herajets}.
This method is very similar to the one based
on event shapes from {\epem} annihilations and
has similar sources of systematic uncertainty.
The present accuracy of the result for
{\amsmz} from DIS jet rates is about~$8\,\%$.
It is not likely that this method will yield a result
for {\amsmz} with precision better than about $5\,\%$
unless a next-to-next-to leading order QCD calculation becomes available.
The overall situation for this measurement is
similar to that for jet rates from {\epem} collisions
and we will not discuss it further.
 
From the above list,
it is seen that the most precise DIS results
for {\alphas} are obtained from the {\qsq} evolution
of {\ftwo} at high {\xf} (item~1),
the {\qsq} evolution of {\xfthree} (item~3),
the Bjorken sum rule (item~5),
and --~with somewhat less precision at present~--
the GLS sum rule (item~4).
It is of note that all three structure functions
{\ftwo}, {\fthree} and {\gone} contribute at least
one measurement with 4-$5\,\%$ accuracy,
illustrating the strength of the complementarity
offered by the unpolarized charged lepton,
neutrino, and polarized charged lepton programs.
The results utilizing the {\qsq} evolution
of {\ftwo} at low {\xf} (item~2) and the shape of {\ftwo}
(item~6) are less accurate.
Method~2 is unlikely to provide a precise result
for {\alphas} in the future,
since the {\qsq} evolution
of the singlet component at small {\xf}
depends on the gluon distribution function {\gdist},
as mentioned above:
this situation will not change for future data sets.
If data are collected using different nuclear
targets so that the singlet and non-singlet components
of {\ftwo} can be separated,
the evolution of the non-singlet component of
{\ftwo} at relatively small {\xf} values could still be
a viable method for an accurate {\alphas} result:
this was not found to be the case in~\cite{bib-nmc93},
however, which included such an analysis.
It is more difficult to assess the future status
of the {\alphas} result based on method~6
since this method has only recently been proposed.
This method is based on the asymptotic behavior
of the QCD resumed prediction for {\ftwo}
at large {\qsq} and small {\xf} and
has been applied to HERA data.
The dominant uncertainty
arises from the ambiguity in the choice of the
renormalization and factorization scales~\cite{bib-ball95}.
This suggests that a reduction in the uncertainty
of {\amsmz} below the $5\,\%$ level will
require the inclusion of subleading terms,
the prospects for which are unknown.
Further theoretical understanding of this method
will probably be required before it can be used to
accurately measure~{\alphas}.
We will not consider this method further.
The remaining discussion on the prospects
for a precise {\alphas} measurement from DIS
therefore concentrates on items 1, 3, 4 and~5
in the above list.
 
\subsection{Future Prospects for a Precise
     {\alphas} Measurement from DIS}
 
The future facilities which we consider for the
purpose of evaluating the potential for a $1\,\%$
measurement of {\amsmz} from DIS
are the following:
\begin{enumerate}
  \item HERA with a luminosity upgrade,
    able to deliver data samples of about
    150~pb$^{-1}$ per year,
    yielding a total data sample for the
    HERA experiments of 500-1000~pb$^{-1}$;
  \item an electron--hadron collider utilizing the LHC,
    referred to as ``LEP$\times$LHC'' in the following;
  \item a $\nu$ ({\nubar}) beam
    from the Tevatron with upgraded luminosity,
    i.e.~Tevatron ``Run~2'' and TeV33,
    available for fixed target experiments;
    it should be emphasized that the prospective
    fixed target neutrino facility
    under consideration here would make use of the
    {\it full energy} Tevatron beam;
  \item a $\nu$ ({\nubar}) beam from the LHC,
    available for fixed target experiments;
  \item future experiments measuring the polarized
    structure function~{\gone}.
\end{enumerate}
The facilities listed above have often been
presented as natural extensions to
the HERA, Tevatron, and LHC programs,
with the possible exception of a fixed target
facility at the LHC.
It is not clear whether it is feasible to incorporate
a fixed target neutrino facility into the LHC program.
 
We next discuss each of items 1, 3, 4 and~5
from section~\ref{sec-dis-methods}
in the context of these future facilities.
 
\subsubsection*{{\qsq} Evolution of {\ftwo} at High {\xf} }
 
A measurement of {\alphas} from the {\qsq} evolution
of {\ftwo} is best performed using
charged lepton scattering on unpolarized targets
(for a review of this method,
see~\cite{bib-virchaux92}).
The relevant future facilities for this measurement
are an upgraded HERA and LEP$\times$LHC.
 
An {\alphas} measurement based on the evolution
of {\ftwo} is not possible using the current HERA
data sample because of the scarcity of data at
{\xf} values above about 0.20.
HERA has kinematic access to {\xf} values
up to about 0.50 for {\qsq} values
greater than 10$^3$~GeV$^2$, however.
For the purposes of an {\alphas} measurement,
HERA data at {\qsq}$\,\sim\,$10$^4$~GeV$^2$
and {\xf}$\,\sim\,$0.50 are interesting because
the large {\xf} value ensures suppression
of the contribution from the gluon distribution function,
while the {\qsq} value is similar to that in $Z^0$ decays:
this offers the opportunity
for a direct comparison of the DIS results
with those from {\epem} $Z^0$ experiments.
The region {\qsq}$\,\sim\,$10$^4$~GeV$^2$,
{\xf}$\,\sim\,$0.50
is near the kinematic limit of HERA,
making it likely that data samples of about
1000~pb$^{-1}$ will be necessary for an accurate
{\alphas} measurement based on the evolution
of {\ftwo}.
Furthermore,
weak effects due to $Z^0$ exchange contribute
to the neutral current cross sections for
such {\qsq} values.
It will be necessary to combine electron--proton
and positron-proton data in order to correct for
the weak interference terms,
leading to additional possibilities for systematic error.
It has been estimated~\cite{bib-klein}
that an uncertainty on {\amsmz} of about 0.002
might ultimately be achieved from measurements of
the evolution of {\ftwo} at HERA,
implying a precision of 1.5-$2.0\,\%$.
Such a precision may require a combination of HERA
and fixed target results for {\ftwo}, however~\cite{bib-klein}.
 
Another possibility which has been envisioned
is to operate HERA with electron--deuteron collisions.
Comparison of the electron--proton and
electron--deuteron data would allow the
singlet component of {\ftwo} to be extracted.
A recent study~\cite{bib-klein} implies that this
method could provide an improvement of
about $25\,\%$ in the uncertainty of {\alphas},
relative to what can be achieved using the
electron--proton data alone.
 
The comments made above emphasize the relevance
of considering electron--proton, positron--proton and
electron--deuteron options for LEP$\times$LHC.
A detailed estimate of the {\alphas} precision
achievable using LEP$\times$LHC has not yet been made.
Assuming that there is not a great difference
between the systematic sources of uncertainty
at HERA and LEP$\times$LHC,
it may be presumed that an {\alphas} measurement
with a precision on the order of $2\,\%$ is also possible
at this latter facility.
We note that LEP$\times$LHC offers the possibility
for an accurate determination of {\alphas}
in the {\qsq} range of 2-3$\cdot10^5$~GeV$^2$,
i.e. the same {\qsq} range as a 500~GeV
{\epem} collider.
 
We therefore conclude that an {\alphas} result
with a precision of about $2\,\%$ is a possibility for HERA
at a {\qsq} value of about 10$^4$~GeV$^2$.
Extrapolating to LEP$\times$LHC,
a measurement of similar accuracy may be possible
at a {\qsq} value of about 2-3$\cdot10^5$~GeV$^2$.
 
\subsubsection*{{\qsq} Evolution of {\xfthree}}
 
The {\qsq} evolution of {\xfthree} offers an
advantageous method to measure {\alphas}
because it is independent of the gluon
distribution function {\gdist} over the
entire {\xf} range, as noted above.
Measurements of {\xfthree} are best obtained
using the difference between the
$\nu$ and {\nubar} cross sections for scattering
on unpolarized targets~\cite{bib-virchaux92}.
These measurements require a fixed target program
in order to collect adequate collision statistics.
There is an active experiment at Fermilab
(the NuTeV Collaboration),
which is expected to improve the precision
on {\amsmz} to about $2.5\,\%$
within the next few years,
using this method~\cite{bib-nutev}.
This is likely to become one of the world's most
precise measurements of {\alphas}
and to remain so for some time.
The uncertainties on the NuTeV result are roughly evenly divided
between statistical and systematic sources, with the systematic
uncertainty dominated by imprecise knowledge of the neutrino
beam flux and of the calorimeter energy scale.
 
To improve the precision of the {\alphas} result
from this technique yet further,
higher statistics from tagged neutrino beams
will be necessary
(tagged neutrino beams allow an event-by-event determination
of the incident neutrino energy, as well as {\it apriori}
knowledge of whether the interaction was caused by a neutrino
or an antineutrino).
The future facilities which could potentially
provide beams for a precise {\xfthree} measurement
of {\alphas} are therefore the primary Tevatron beam with
an upgraded luminosity, such as TeV33,
and the LHC.
Given the good result for {\alphas}
which is anticipated from the ongoing experiment,
mentioned above,
and given the improvement in accuracy expected from higher
statistics and the introduction of event-by-event neutrino tagging,
it is plausible that this method can
provide an {\alphas} measurement with $1\,\%$ precision.
A study of the precision attainable
at a LHC fixed target experiment
has not yet been performed, however.
 
In conclusion,
the method based on the
{\qsq} evolution of {\xfthree} is a strong candidate
to provide a $1\,\%$ measurement of {\amsmz},
assuming that fixed target programs with tagged neutrino beams
are available at either TeV33 or the LHC.
We note that the necessary matrix elements
are already available at~NLO \cite{nnlo}. 
Since the $\beta$-function is known to three loops, 
all that is needed for a full NNLO analysis of {\alphas}
using the {\xfthree} method are the splitting functions 
calculated at NNLO.
It is reasonable to expect that this result will
become available and that the theoretical uncertainty
will be below~1\%.
 
\subsubsection*{GLS Sum Rule}
 
The situation regarding the
Gross, Llewellyn-Smith (GLS)
sum rule~\cite{bib-gls}
is similar to that discussed above
for the {\qsq} evolution of {\xfthree}
since both methods rely on the {\fthree} structure
function measured in neutrino fixed target experiments.
The GLS sum rule is based on the integral
\begin{equation}
   \label{eq-gls}
   \int_0^1 \left[{xF_3(x,Q^2)}\right]
       \frac{\mathrm{d}x}{x}
         \; ,
\end{equation}
the QCD prediction for which has been calculated
to $\cal O$(\alphas$^3$)
(the next-to-next-to leading order
in {\alphas}).
This is one of the few quantities calculated to such
a high order in QCD perturbation theory.
The integral (\ref{eq-gls}) is evaluated experimentally
using fairly low values of {\qsq},
which allows small values of {\xf}.
(The small {\xf} region is particularly important
because of the 1/{\xf} weighting in~(\ref{eq-gls}).)
The {\qsq} value for present
experiments~\cite{bib-ccfr93} is about 3~GeV$^2$.
Because of the low {\qsq} value,
higher twist corrections are important.
Furthermore,
it is necessary to extrapolate
into the unmeasured region at low~{\xf}.
 
Like the result for {\alphas} based on the evolution
of {\xfthree},
the current precision of the {\alphas}
measurement from the GLS sum rule
is partially statistics-limited.
There are several sources of experimental systematic
uncertainty which are relevant for
the GLS result and which are not relevant for the
{\xfthree} evolution result, however:
the measurement of the absolute cross sections
for both the $\nu$ and {\nubar} beams,
the extrapolation into the low {\xf} region,
and higher twist corrections.
The NuTeV experiment
expects to attain a precision of about
$3\,\%$ for {\alphas} using the GLS sum rule.
At TeV33 and the LHC,
higher statistics, larger {\qsq} values
(reducing the uncertainty
from higher twists)
and better low {\xf} reach
(reducing the extrapolation uncertainty)
should yield smaller statistical and
systematic errors,
making a measurement of {\amsmz} with a
precision of $1.5\,\%$ a possibility.
 
In conclusion,
the GLS sum rule provides an independent method
to determine {\alphas} from a
neutrino fixed target experiment,
using the structure function~{\fthree}.
Systematic uncertainties should be reduced at the higher
{\qsq} values offered by TeV33 or the LHC,
relative to the current experiments,
making an {\alphas} measurement with a precision
of about $1.5\,\%$ feasible.
 
\subsubsection*{Bjorken Sum Rule}
 
Lastly,
we consider the determination of {\alphas}
using the Bjorken sum rule~\cite{bib-bjorken}.
The Bjorken sum rule is based on the quantity
\begin{equation}
   \label{eq-bj}
   \int_0^1 \left[{g_1^p(x,Q^2)-g_1^n(x,Q^2)}\right]
   \mathrm{d}x
   \; ,
\end{equation}
where g$_1^{{p}}$ and g$_1^{{n}}$
are the {\gone} structure functions for proton
and neutron targets, respectively.
The Bjorken sum rule method for determining {\alphas}
differs from the others discussed here
in that it is based on polarized cross sections.
The method resembles the GLS sum rule
technique, however,
because it is based on an integral over {\xf}
of a structure function measured
at fixed {\qsq},
utilizes low {\qsq} measurements
(for current experiments,
the {\qsq} value is about 2~GeV$^2$),
requires extrapolation into the unmeasured
{\xf} regions,
and has a QCD prediction available at the
next-to-next-to leading order.
 
Like the method based on the shape of {\ftwo}
(item~6 in section~\ref{sec-dis-methods}),
it is somewhat difficult to assess the future status
of the {\alphas} result obtainable from the Bjorken
sum rule because it is only recently that this method
has been used to determine~{\alphas}.
The current result
(about $4\,\%$ accuracy~\cite{bib-ellis95})
is quite precise by current standards, however.
Given that additional polarized structure function
data are currently being collected at CERN, DESY and SLAC,
and that additional experiments are being planned,
it can be anticipated that a reduction in the uncertainty
in {\alphas} from this method will be possible.
Many sources of systematic uncertainty
(higher twists, extrapolation into the unmeasured {\xf} region,
measurement of the absolute cross sections)
are common between this method and the GLS one.
The Bjorken sum rule measurement is complicated
by its reliance on polarized targets and probes, however,
and thus has sources of systematic uncertainty which are
not present for the GLS measurement.
Therefore,
we presume that the ultimate accuracy for {\amsmz}
achievable from the Bjorken sum rule
for experiments currently running or being planned
lies between the current
precision ($4\,\%$) and that which we estimate will be
achievable from the GLS sum rule ($1.5\,\%$).
Thus, an estimate of about $2.5\,\%$ precision seems justifiable.
 
Although no study has been done at this point,
we wish to emphasize
that some of the systematics which degrade the accuracy
of the Bjorken sum rule measurement of {\alphas},
including those due to higher
twist and the low {\xf} extrapolation,
may improve with a high statistics,
high energy beam.
Such a beam would be available if a high energy
{\epem} collider were constructed with longitudinal
polarization and a fixed-target capability.
In this way, it is plausible that the Bjorken sum rule measurement
could be more accurate than the estimate given above.

\subsection{Conclusion for a Precise {\alphas} Result from DIS}
 
\begin{table}[t]
\caption{
The estimated precision for {\amsmz} attainable
at future DIS experiments.
}
\centering
\setlength{\tabcolsep}{9pt}
\let\mcol=\multicolumn
\begin{tabular}{|l|c|l|}
  \hline  &&\\[-10pt]
\mcol{1}{|c|}{{\em Method}} & {\em Precision} &
      \mcol{1}{c|}{{\em Facility}} \\[2pt]
  \hline
  \hline  &&\\[-8pt]
    {\qsq} evolution of {\ftwo}   & $2\,\%$   & HERA, LEP$\times$LHC \\
    at high {\xf}                 &       & \\[2mm]
    {\qsq} evolution of {\fthree} & $1\,\%$   & TeV33 fixed target,  \\
                                  &       & LHC fixed target\\[2mm]
    GLS sum rule                  & $1.5\,\%$ & TeV33 fixed target,  \\
                                  &       & LHC fixed target\\[2mm]
    Bjorken sum rule             & $2.5\,\%$ &  Future polarized \\
                                  &       &  DIS experiments \\[2pt]
  \hline
\end{tabular}
\label{tab-dissummary}
\end{table}
 
Table~\ref{tab-dissummary} summarizes our estimates
of the precision which might be attainable for
{\amsmz} from DIS experiments at future colliders.
These estimates are mostly based on extrapolations
from current experiments rather than on detailed
studies of future facilities.
The best prospect for a $1\,\%$ measurement of {\amsmz}
from a DIS experiment is from a fixed target neutrino
facility at a hadron collider with
high flux, tagged $\nu$ and {\nubar} beams.
The most promising measurement technique is the
observation of the {\qsq} evolution of~{\xfthree}.
 
We again emphasize, however, the importance of accurate
{\alphas} measurements at widely different {\qsq} values.
The DIS results based on {\xfthree} offer the possibility
for a precise  measurement of {\alphas} in the {\qsq} range
from about 5 to 20~GeV$^2$.
Those based on the high {\xf} region of {\ftwo} offer
the possibility for an accurate measurement at a much
larger {\qsq} value, up to about 10$^5$~GeV$^2$ for
LEP$\times$LHC.

\section{The Hadron Spectrum}
\label{sec:IM}

Lattice QCD is, so far, the only systematic, first principles approach
to nonperturbative QCD. The experimentally observed hadron spectrum,
like the high-energy observables discussed in the other sections,
provide us with information on the free parameter of QCD, {\alphas}.
Determinations of {\alphas} from the experimentally observed hadron spectrum,
based on lattice QCD, are thus complementary to determinations which are
based on perturbative QCD.
 
While an introduction to lattice QCD is beyond the scope of this report
(see \cite{intro} for pedagogical introductions and reviews), we shall, in
the following, outline the strategy for determinations of {\alphas} based
on lattice QCD. In general, determinations of {\alphas} can be divided into
three steps:
 
The first step is always an experimental measurement.
In $\alpha_s$ determinations based on perturbative QCD this might be a
cross section or (ratio of) decay rates. In determinations based on lattice
QCD this is usually a hadron mass or mass splitting, for example the mass
of the $\rho$ meson, or a better choice, spin-averaged splittings in the
charmonium and bottomonium systems. In ``lattice language'' this step is often
referred to as ``setting the scale'' (see section~\ref{sec:scale}).
 
The second step involves a choice of renormalization
scheme. In perturbative QCD the standard choice is the $\ms$ scheme.
With lattice QCD a nonperturbative
scheme may be chosen, and there are many candidates. In order to compare
with perturbative QCD, any such scheme should be accessible to perturbative
calculations (without excessive effort).
 
Finally, the third step is an assessment of the
experimental and theoretical errors associated with the strong coupling
determination. This is of course the most important (and sometimes also the
most controversial) step as it allows us to distinguish and weight different
determinations. The experimental errors on hadron masses are negligibly
small in lattice determinations of $\alpha_s$ at this point.
The theoretical errors that are part of $\alpha_s$ determinations based on
perturbative QCD include higher order terms in the truncated perturbative
series and the associated dependence on the renormalization scale, and
hadronization or other generic nonperturbative effects. In lattice QCD
the theoretical errors include (but are not limited to) discretization
errors (due to the finite lattice spacing, $a\neq 0$), finite volume
effects, and errors associated with the partial or total omission of sea
quarks.
 
The consideration of systematic uncertainties should guide us towards
quantities where these uncertainties are controlled, for a reliable
determination of $\alpha_s$.
As has been argued by Lepage \cite{lepage}, quarkonia are among
the easiest systems to study with lattice QCD, since systematic
errors can be analyzed easily  with potential models if not by
brute force.
 
Finite-volume errors are much easier to control for
quarkonia than for light hadrons, since quarkonia are smaller.
Lattice-spacing errors, on the other hand, can be larger
for quarkonia and need to be considered.
This error can be controlled by studying the lattice spacing dependence
of physical quantities (in physical units). The lattice spacing is
reduced (while keeping the physical volume of the lattice fixed) until
the error is under sufficient control.
The source of the lattice spacing dependence are the discretizations used
in the lattice lagrangian (or action).
Thus, an alternative to reducing the lattice spacing in order to control
this systematic error is the use of better discretizations. This procedure
is generally referred to as improving the action.
For quarkonia, the size of lattice-spacing errors in a numerical
simulation can be {\em anticipated} by calculating expectation
values of the corresponding operators using potential-model wave
functions. They are therefore ideal systems to test and establish
improvement techniques.
 
A lot of the work of phenomenological relevance is done in
what is generally referred to as the ``quenched''
(and sometimes as the ``valence'') approximation.
In this approximation gluons are not allowed to split into
quark - anti-quark pairs (sea quarks). This introduces a systematic
error into the calculation.
However, for quarkonia, a number of calculations now exist which
partially include the effect of sea quarks, thereby significantly
reducing this systematic error. This is further discussed in
sections~\ref{sec:scale} and \ref{sec:sea}.
 
\subsection{Determination of the Lattice Spacing and the Quarkonium Spectrum}
   \label{sec:scale}

\begin{figure}[htbp]
\centering
\centerline{\psfig{file=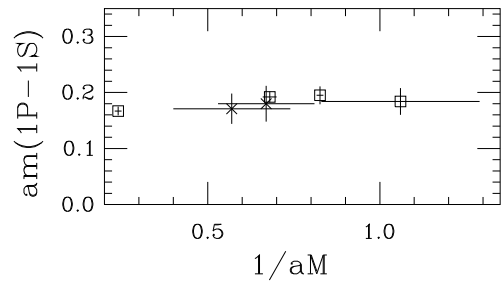,width=6.5cm}}
\caption[xxx]{The 1P-1S splitting as a function of the 1S mass
   (statistical errors only) from Ref.~\cite{us};
   $\square$: ${\cal O}(a^2)$ errors; $\times$: ${\cal O}(a)$ errors.}
\label{fig:1p1s}
\end{figure}

The experimental input to the strong coupling determination is
a mass or mass splitting, from which by comparison with the
corresponding lattice quantity the lattice spacing, $a$, is determined in
physical units.
For this purpose, one should identify quantities that are insensitive
to lattice errors. In quarkonia, spin-averaged splittings are good
candidates. The experimentally observed 1P-1S and 2S-1S splittings
depend only mildly on the quark mass (for masses between $m_b$ and $m_c$).
Figure~\ref{fig:1p1s}
shows the observed mass dependence of the 1P-1S
 splitting
in a lattice QCD calculation. The comparison between results from
different lattice actions illustrates that
higher-order lattice-spacing errors for these splittings
are small \cite{nrqcd_als,us}.
 
\break

Two different formulations for fermions have been used in lattice
calculations of the quarkonia spectra.
In the non-relativistic limit the QCD action can be written as
an expansion in powers of $v^2$ (or $1/m$), where $v$ is the
velocity of the heavy quark inside the bound state \cite{eff};
Henceforth, this approach shall be referred to as NRQCD. Lepage and
collaborators \cite{nrqcd_thy} have adapted this formalism to
the lattice regulator. Several groups have performed numerical
calculations of quarkonia in this approach.
In \mbox{Refs.~\cite{nrqcd_spec,nrqcd_cc}} the
NRQCD action is used to calculate
the $\bb$ and $\cc$ spectra, including
terms up to ${\cal O} (mv^4)$ and ${\cal O}(a^2)$.
In addition to calculations in the quenched approximation,
this group is also using gauge configurations that include
two flavors of sea quarks with mass $m_q \sim \frac{1}{2} m_s$
to calculate the $\bb$ spectrum \cite{nrqcd_als,shige}.
The leading order NRQCD action is used in Ref.~\cite{uknrqcd}
for a calculation of the $\bb$ spectrum in the quenched approximation.
 
The Fermilab group \cite{us_thy} developed a generalization of previous
approaches, which encompasses the non-relativistic
limit for heavy quarks as well as Wilson's relativistic action
for light quarks. Lattice-spacing errors are analyzed for quarks with
arbitrary mass. Ref.~\cite{us} uses this approach to calculate
the $\bb$ and $\cc$ spectra in the quenched approximation. The authors
considered the effect of reducing lattice-spacing errors from ${\cal O}(a)$
to ${\cal O}(a^2)$.
The SCRI collaboration \cite{sloan} is also using this approach
for a calculation of the $\bb$ spectrum using the same gauge configurations
as the NRQCD collaboration with $n_f = 2$ and an improved fermion action
(with ${\cal O}(a^2)$ errors).
 
\begin{figure}[htbp]
\centering
\centerline{\psfig{file=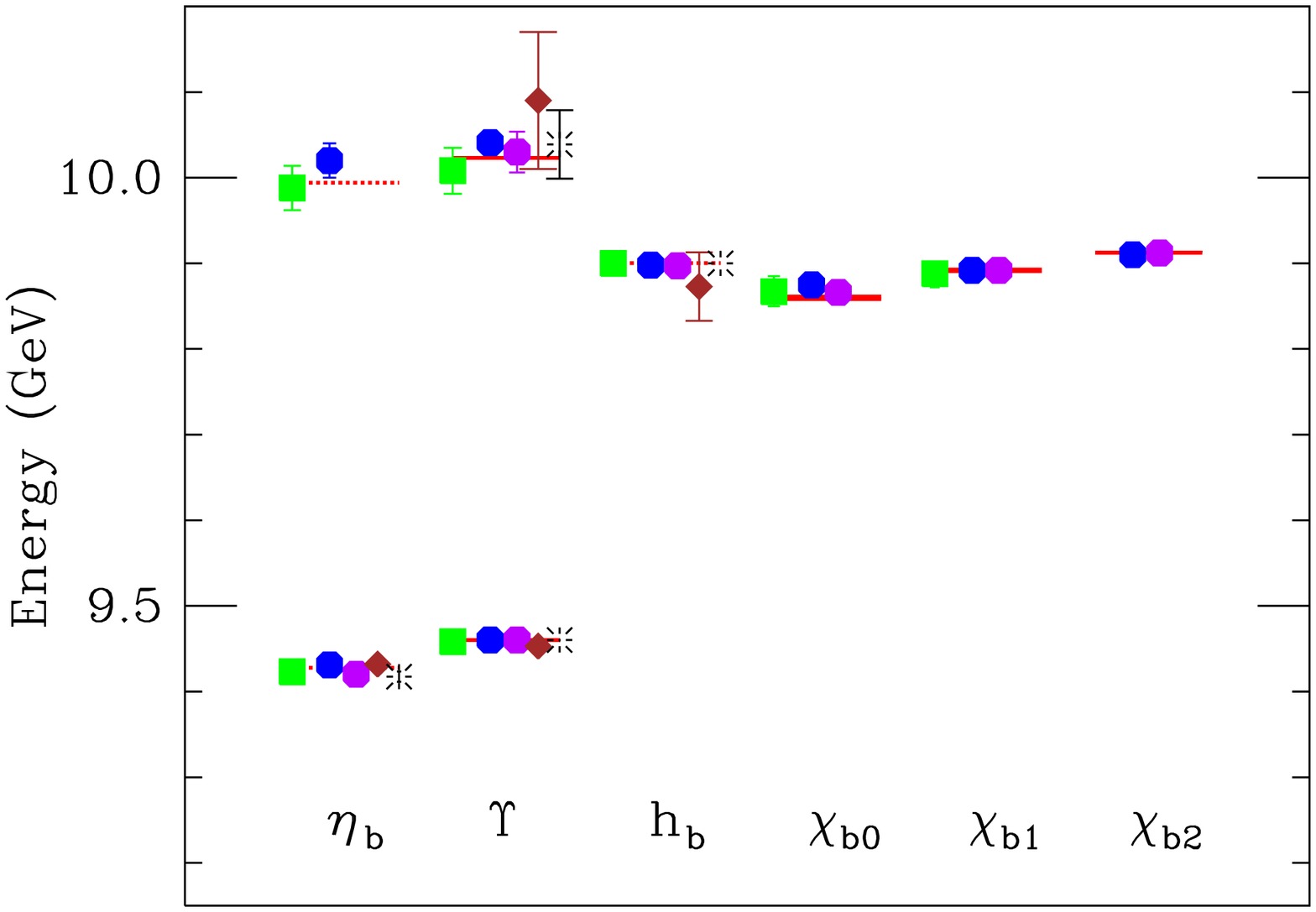,width=8.5cm}}
\caption[xxx]{A comparison of lattice QCD results for the $\bb$ spectrum
   (statistical errors only).
 --: Experiment; $\square$: FNAL~\protect\cite{us};
 $\circ$: NRQCD ($n_f=0$)~\protect\cite{nrqcd_spec};
 $\bullet$: NRQCD ($n_f=2$)~\protect\cite{nrqcd_als};
 $\Diamond$: UK(NR)QCD~\protect\cite{uknrqcd}; $*$: SCRI~\cite{sloan}.}
   \label{fig:bb}
\vskip2pc
\centering
\centerline{\psfig{file=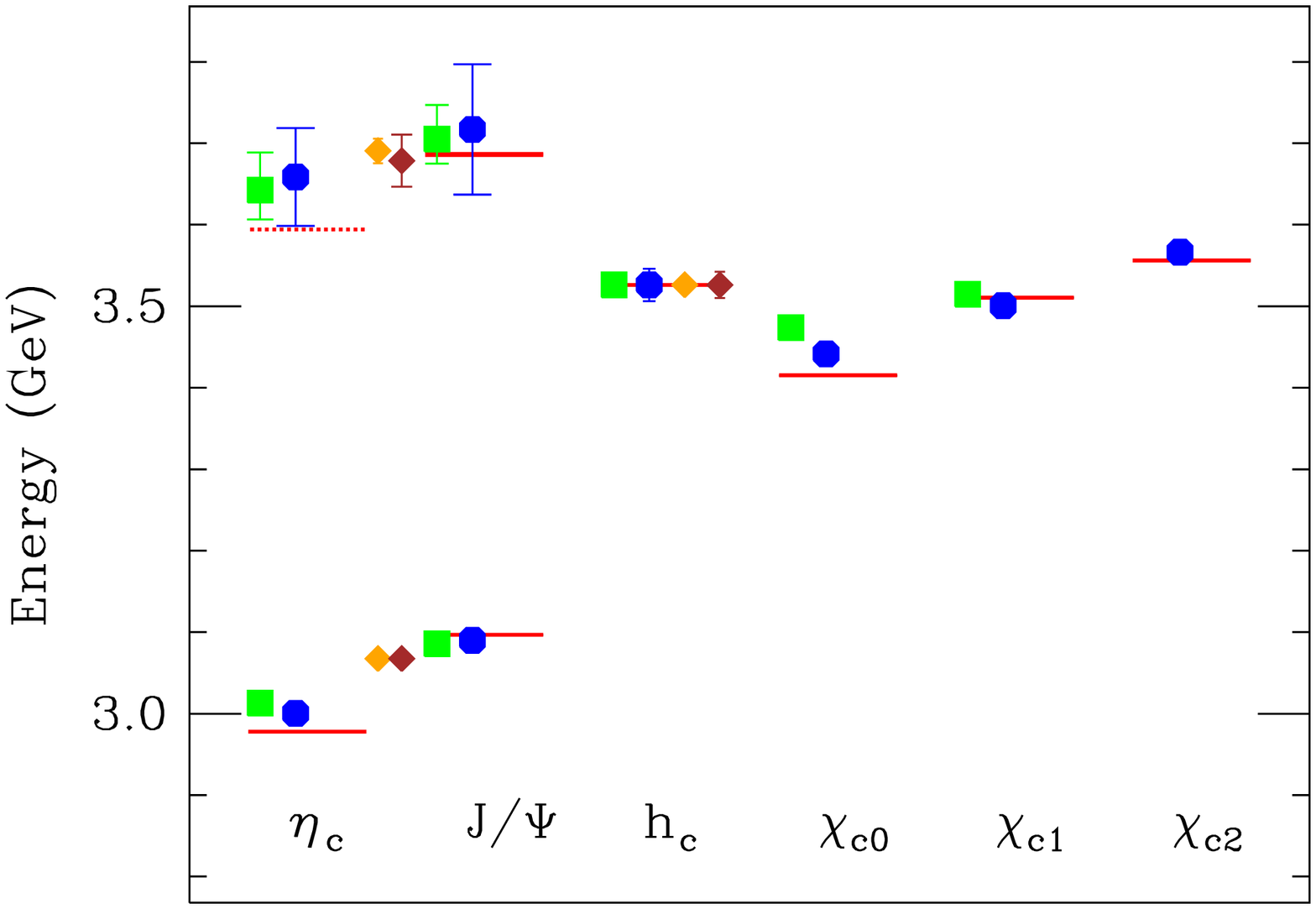,width=8.5cm}}
\caption[xxx]{A comparison of lattice QCD results for the $\cc$ spectrum
   (statistical errors only).
   --: Experiment; $\square$: FNAL~\protect\cite{us}; $\circ$:
   NRQCD ($n_f=0$)~\protect\cite{nrqcd_cc}; 
$\Diamond$: ADHLM~\protect\cite{adhlm}.}
 \label{fig:cc}
\end{figure}

All but one group use gauge configurations generated with the Wilson
action, leaving ${\cal O}(a^2)$ lattice-spacing errors in the
results. The lattice spacings, in this case, are in the range
$a \simeq 0.05 - 0.2$ fm.
Ref.~\cite{adhlm} uses an improved gauge action together with
a non-relativistic quark action improved to the same order
(but without spin-dependent terms) on coarse ($a \simeq 0.4 - 0.24$ fm)
lattices.
The results for the $\bb$ and $\cc$ spectra from all groups are
summarized in Figures~\ref{fig:bb}~and~\ref{fig:cc}.
 
The agreement between the experimentally-observed spectrum and
lattice QCD calculations is impressive. As indicated in the
preceding paragraphs, the lattice artifacts are different for all groups.
Figures~\ref{fig:bb}~and~\ref{fig:cc} therefore emphasize the level of
control over systematic errors.
 
Results with two flavors of degenerate sea quarks
have now become available from a number of
groups \cite{nrqcd_als,kek_als,cdhw,shige}, with lattice-spacing
and finite-volume errors similar to the quenched calculations,
significantly reducing this systematic error.
However, several systematic effects associated with the
inclusion of sea quarks still need to be studied further.
They include the dependence of the quarkonium
spectrum on the number of flavors of sea quarks, and
the sea-quark action (staggered vs. Wilson). The inclusion
of sea quarks with realistic light-quark masses is very
difficult and can, at present, only be done by extrapolation from
$m_q\simeq 0.3-0.5 m_s$ to $m_{u,d}$.
However, the dependence of the quarkonium splittings on the sea quark
masses can be analyzed with chiral perturbation theory \cite{gr} to
guide the extrapolation.

\subsection{Definition of a Renormalized Coupling}

Within the framework of lattice QCD the conversion from the bare
to a renormalized coupling can, in principle, be made
nonperturbatively. In the definition of a renormalized coupling,
systematic uncertainties should be controllable, and at short
distances, its (perturbative) relation
to other conventional definitions calculable.
For example, the renormalized coupling, $\al_V$, can
be defined from the nonperturbatively computed heavy-quark potential
\cite{schill}.
An elegant approach has been developed in Ref.~\cite{luesch}, where
a renormalized coupling is defined nonperturbatively through
the Schr\"{o}dinger functional. The authors compute the evolution
of the coupling nonperturbatively using a finite size
scaling technique, which allows them to vary the momentum
scales by an order of magnitude.
The same technique has also been applied to the renormalized coupling
defined from twisted Polyakov loops \cite{alpha}. The numerical
calculations include only gluons at the moment. However, the inclusion of
fermions is possible. Once such simulations become available they should
yield very accurate information on {\alphas} and its evolution.
A renormalized coupling can also be defined from the three-gluon
vertex, suitably evaluated on the lattice \cite{parr}.
 
An alternative is to define a renormalized coupling through short
distance lattice quantities, like small Wilson loops or Creutz
ratios which can be calculated perturbatively and by numerical simulation. For
example, the coupling defined from the plaquette (the smallest Wilson loop
on the lattice),
$\al_{\square} = - 3 \ln{ \langle \Tr \, U_{\square} \rangle } / 4 \pi$,
can be expressed as \cite{lm}:
\begin{equation} \label{eq:lm19}
 \al_{\square} = \al_P (q) [ 1 - (1.19  + 0.07 \, n_f) \al_P (q) ]
\end{equation}
at $q=3.41/a$, close to the ultraviolet cut-off.
The coupling $\al_P$ is chosen such that it equals $\al_V$ at one-loop:
\be
  \al_P (q) = \al_V (q) + {\cal O} (\al_V^3) \;.
\ee
$\al_P$ is related to the more commonly used $\ms$ coupling by
\be    \label{eq:vms}
\al_{\ms} (Q) = \al_P (e^{5/6} Q) \left(1 + \nicefrac{2}{\pi} \al_P
              + c_2 (n_f) \al_P^2 + \ldots \right) \;.
\ee
The size of higher-order corrections associated with the above
defined coupling constants can be tested by comparing
perturbative predictions for short-distance lattice quantities
with nonperturbative results \cite{lm}.
The comparison of the nonperturbatively calculated coupling of
Ref.~\cite{luesch} with the perturbative predictions for this coupling
using Eq.~(\ref{eq:lm19}) is an additional consistency test.
 
The relation of $\al_P$ to $\al_{\ms}$ ,Eq.~(\ref{eq:vms}), has
recently been calculated to two loops \cite{alles,lw-2l} in the
quenched approximation (no sea quarks, $n_f = 0$):
\be
  c_2 (n_f = 0) = 0.96 \;.
\ee
This term shifts {\afmsmz} by $+0.002$. Because of the unknown $n_f$
dependence in the two-loop term, $c_2$, the perturbative uncertainty is
still $\pm 0.002$ (at {\mz}). The extension of the two-loop calculation
to $n_f \neq 0$ will reduce this uncertainty to well
below $1\,\%$ for {\afmsmz}.
 
\subsection{Sea Quark Effects}   \label{sec:sea}

Calculations that properly include all sea-quark effects
do not yet exist.
If we want to make contact with the ``real world'', these effects
have to be estimated phenomenologically or extrapolated away.
 
The phenomenological correction necessary to account for
the sea-quark effects omitted in calculations of quarkonia
that use the quenched approximation gives rise to the dominant
systematic error in this calculation \cite{prl,nrqcd_l93}.
By demanding that, say, the spin-averaged 1P-1S splitting calculated on
the lattice reproduce the experimentally observed one (which
sets the lattice spacing, $a^{-1}$, in physical units), the effective
coupling of the quenched potential is in effect matched to the
coupling of the effective three flavor potential at the typical
momentum scale of the quarkonium states in question. The difference
in the evolution of the zero flavor and 3,4 flavor couplings
from the effective low-energy scale to the ultraviolet cut-off,
where {\alphas} is determined, is the perturbative estimate
of the correction.
 
For comparison with other determinations of {\alphas}, the $\ms$
coupling can be evolved to the $Z^0$ mass scale. An average
of Refs.~\cite{prl,nrqcd_l93} yields for {\alphas} from calculations
in the quenched approximation:
\be          \label{eq:nf0}
   \al^{(5)}_{\ms} ({M}_{Z}) = 0.110 \pm 0.006 \;.
\ee
 
The phenomenological correction described in the previous paragraph
has been tested from first principles in Ref.~\cite{kek_als}.
The 2-loop evolution of $n_f = 0$ and $n_f = 2$ $\ms$ couplings
-- extracted from calculations of the $\cc$ spectrum using the
Wilson action in the quenched approximation and with two flavors of
sea quarks respectively -- to the low-energy scale gives consistent
results. After correcting the two flavor result to $n_f = 3$ in the
same manner as before and evolving $\al_{\ms}$ to the $Z^0$ mass,
they find \cite{kek_als}
\be
 \al^{(5)}_{\ms} ({M}_{Z}) = 0.111 \pm 0.005
\ee
in good agreement with the previous result in Eq.~(\ref{eq:nf0}).
The total error is now dominated by the rather large
statistical errors and the perturbative uncertainty.
 
The most accurate result to date comes from the NRQCD
collaboration \cite{nrqcd_als,shige}.
They use results for {\alphas} from the $\bb$
spectrum with 0 and two flavors of sea quarks to extrapolate
the inverse coupling to the physical three flavor case directly at
the ultraviolet momentum, $q = 3.41/a$. They obtain a result
consistent with the old procedure.
Recently, they have begun to study the dependence of {\alphas} on
the masses of the sea quarks. Their preliminary result is:
\be   \label{eq:v3}
    \al_P^{(3)} (8.2 \, \GeV) = 0.195 \pm 0.003 \pm 0.001 \pm 0.004 \;.
\ee
The first error is statistics, the second error an estimate of
residual cut-off effects and the third (dominant) error is due to the
quark mass dependence.
The conversion to $\ms$ (including the 2-loop term in Eq.~(\ref{eq:vms})
and evolution to the $Z^0$ mass then gives:
\be        \label{eq:ms5}
\al^{(5)}_{\ms}({M}_{Z}) = 0.118 \pm 0.003 \;,
\ee
where the error now also includes the perturbative uncertainty from
eq.~(\ref{eq:vms}). A similar analysis is performed in Ref.~\cite{cdhw}
on the same gauge configurations but using the Wilson action for
a calculation of the $\cc$ spectrum. The result for the coupling
is consistent with Refs.~\cite{nrqcd_als,kek_als}.
 
The preliminary calculation of the SCRI collaboration \cite{sloan}
($n_f=2$) can be combined with the result of Ref.~\cite{us}. Using the
same analysis as in Ref.~\cite{nrqcd_als} gives \cite{shige}
\be \label{eq:ms5n}
\al^{(5)}_{\ms}({M}_{Z}) = 0.116 \pm 0.003 \;\;,
\ee
nicely consistent with Eq.~(\ref{eq:ms5}).
Clearly, more work is needed to confirm the results of Eqs.~(\ref{eq:ms5})
and (\ref{eq:ms5n}), especially in calculations with heavy quark actions
based on Ref.~\cite{us_thy}.
In particular, the systematic errors associated with the
inclusion of sea quarks into the simulation have to be checked,
as outlined in section~\ref{sec:scale}.
 
\subsection{Conclusions} \label{sec:con}

Phenomenological corrections are a necessary evil that enter most
coupling constant determinations.
In contrast, lattice QCD calculations with control over
all sources of systematic error can, at least in principle, yield truly
first-principles
determinations of {\alphas} from the experimentally observed hadron
spectrum.
 
At present, determinations of {\alphas} from the experimentally measured
quarkonia spectra using lattice
QCD are comparable in reliability and accuracy to other determinations
based on perturbative QCD from high energy experiments (see
Figure~1).
The phenomenological corrections for the most important sources
of systematic errors in lattice QCD calculations of quarkonia have already
been replaced by first principles calculations. This has led to a
significant increase in the accuracy of {\alphas} determinations from
quarkonia.
 
Still lacking for a first-principles result is the proper inclusion of
sea quarks. A difficult problem in this context is the inclusion of sea
quarks with physical light quark masses. At present, this can only be
achieved by extrapolation (from $m_q \simeq 0.3 - 0.5 m_s$ to $m_{u,d}$).
Given sufficient numerical results on the light quark mass dependence,
chiral perturbation theory can be used for the extrapolation \cite{gr}.
These calculations can most likely be 
performed with currently available computational
resources leading to first-principles results for the quarkonia spectra.
They should, in turn, yield determinations of
{\amsmz} with a total uncertainty below $1\,\%$.

\section{Electron-Positron Annihilation}
 
The measurement of {\alphas} via hadronic observables in {\epem}
annihilation is a mature subject. Prospects for the accurate
measurement of {\alphas} in high-energy {\epem} annihilation
have been under discussion for some time~\cite{bib-Mor}.
 To assess the potential for an {\alphas} measurement from
 this method, it is useful to examine an experimental
 analysis in detail, to assess those areas in which
 the uncertainties might be reduced in the future.
 For this purpose, we choose a recent comprehensive
 study of {\alphas} published by the
 SLD Collaboration~\cite{bib-sldas}.  Similar studies
 have been published by the LEP
 experiments~\cite{bib-lepas}.  The SLD result
\be
  \al_{\ms} ({M}_{Z}) = 0.1200 \pm 0.0025 \pm 0.0078
\ee
was derived
from the consideration of 15 different infrared-safe
hadronic observables, including various event shape parameters,
jet rates derived with several different jet finding schemes,
and energy-energy correlations. The $\pm 0.0025$ experimental
error received contributions of $\pm 0.0009$ from event statistics,
and $\pm 0.0024$ from detector-related uncertainties. 
The $\pm 0.0078$ theoretical uncertainty resulted from
contributions of $\pm 0.0024$ from uncertainties in the hadronization
process, and $\pm 0.0074$ from missing higher orders in the
perturbative calculation of the 15 observables. Currently, all 15
observables have been calculated to next-to-leading order
in {\alphas}. In addition, for six of the 15 observables,
the leading and sub-leading logarithms have been resummed
and combined with the next-to-leading order calculations.
 
This breakdown of the uncertainty provides a basis for
estimating the accuracy of a similar measurement of {\alphas}
at an electron--positron collider of cms energy $\sqrt{s} = 500$~GeV,
such as the proposed NLC.
Statistically, the SLD measurement was performed with the sample of
37,000 hadronic ({\epem}$\,\rightarrow\,${\qq}) events remaining
in the 1993 SLD event sample after the application of hadronic
event selection cuts.
At a design luminosity of $5 \times 10^{33}$~cm$^{-2}$sec$^{-1}$,
with a Born-level
cross section of $3.1$~pb, an NLC detector would
collect approximately 150,000
{\epem}$\,\rightarrow\,${\qq} events in a ``Snowmass'' year of
$10^7$ seconds. The effects of
initial state radiation and
beamstrahlung, and inefficiencies introduced by event selection
(to be discussed below), reduce this to approximately 25,000
{\epem}$\,\rightarrow\,${\qq} events per year
at $\sqrt{s} \simeq  500$ GeV, adequate for a statistical precision
of $\pm 1\,\%$ on the value of~{\alphas} at that scale. A well
designed NLC Detector calorimeter should permit a substantial
reduction in the $\pm 2\,\%$ detector uncertainty.
 
The determination of {\alphas}
involves the comparison of the hadronic observables
with parton-level perturbative calculations which
depend upon~{\alphas}.
The relationship between the parton-level
calculations and the final state observables is thus obscured
by the fragmentation process. This introduces a correction,
and corresponding uncertainty, which must be applied to the
extracted value of~{\alphas}. 
It is generally expected~\cite{q2epem} that
effects which alter
the relation between the perturbative parton-level calculations
of observables, and the actual hadron-level observables,
scale as an inverse power of the momentum transfer~$Q$, so that
for some observable $O$,
\be
\delta O \equiv O - O_{\rm pert} \;\; \sim \; \frac{a}{Q^n} \;.
\ee
Typically, $n$ equals 1 or 2.
On the other hand, the perturbative evolution of {\alphas} scales
roughly as $(\ln{Q})^{-1}$. Thus, one expects the relative
uncertainty on {\alphas} due to fragmentation effects to
scale as $\ln{Q} / Q$. As a result, the $\sim 10\,\%$ correction
applied
to the value of {\alphas} extracted from hadronic observables
at the $Z^0$ pole is expected to reduce to a $\sim 2\,\%$
correction at $\sqrt{s} = 500$ GeV, with an uncertainty
of~$1\,\%$ or less.

In addition to the fragmentation, the relationship between
the perturbative parton-level calculation and the measured
observables is compromised by missing higher orders in the
perturbative expansion. In the SLD analysis, this uncertainty
was estimated to be $\Delta\alpha_{\ms}(M_Z)=\pm 0.0074$
by varying the renormalization
scale of the perturbative calculation
over a range permitted by consistency with the hadronic data,
and observing the corresponding variation in the extracted
value of~{\alphas}. Current perturbative calculations are
done to order $\alpha_S^2$; thus, uncertainties due to
missing higher orders should scale as $\alpha_S^3$, leading
to an uncertainty of $\pm$0.003--0.004 at $\sqrt{s} = 500$ GeV.
Evolving this back to the benchmark scale
{\qsq}=$M_{Z}^2$ using
the three-loop QCD $\beta$-function yields an uncertainty of
$\pm$0.005--0.006, or 4--5\,\% relative, on the value of {\amsmz}
extracted from hadronic observables at the NLC.
Should
next-to-next-to-leading order perturbative calculations become
available, it should thus be possible to approach the target
uncertainty of $\pm 1\,\%$.
 
A final issue associated with the measurement of {\alphas} in
{\epem} annihilation at large cms energy is that of
identifying a clean sample of
{\epem}$\,\rightarrow\,${\qq} ($q\neq t$) events.
At $\sqrt{s} = 500$ GeV, without event selection cuts,
{\qq} ($q\neq t$) production ($\sigma_{Born} = 3.1$ pb)
has a substantially smaller cross section than
$W^+W^-$ production ($\sigma_{Born} = 7.0$ pb),
as well as significant backgrounds from
$Z^0Z^0$ ($\sigma_{Born} = 0.4$ pb)
and {\toptop} ($\sigma_{Born} = 0.3$ pb) production.
A study performed by the European Linear Collider
QCD Working Group~\cite{euro} identified a set of kinematic cuts which
select an $83\,\%$ pure sample of
{\epem}$\,\rightarrow\,${\qq} ($q\neq t$) events. However,
these cuts substantially impacted the hadronic distributions of
the remaining {\qq} events, leading to large ($\sim 20\,\%$)
corrections to the extracted value of~{\alphas}.
To this end,
in preparation for the Snowmass Workshop,
a Monte Carlo study~\cite{bib-evsel} was undertaken in which
one of the European Working Group kinematic cuts -- the requirement
that at least one hemisphere have a reconstructed invariant mass
of less than $13\,\%$ of the visible energy in the event --
was removed. Instead,
events were
used only if they were produced with the right-handed electron
beam (to remove $W^+W^-$ background),
and if they gave no indication of
the presence of B hadrons in the vertex detector (to eliminate
{\toptop} background).
For an electron beam polarization
of $P_e = 80\,\%$ ($P_e = 90\,\%$), this yielded an $82\,\%$ ($87\,\%$)
pure ``Snowmass'' sample of
{\epem}$\,\rightarrow\,${\qq} ($q\neq t$) events.
A comparison of 3-jet rates between a pure Monte Carlo sample of
{\epem}$\,\rightarrow\,${\qq} ($q\neq t$) events, and
the sample identified by the Snowmass cuts, indicated that
corrections due to the Snowmass event selection are
substantially less than $5\,\%$. Thus, with these cuts, the uncertainty
on {\alphas} due to the event selection process is expected to be
well within the target of $\pm 1\,\%$.
It should be noted that electron beams with $80\,\%$ polarization,
and bunch populations
exceeding that required for the operation of the NLC, are already
in use at the SLAC Linear Collider,
and that polarized running is part of the base-line
proposal for the NLC~\cite{bib-nlcrep}.

As a final note, it has been pointed out~\cite{bib-Sighaw} 
that the high luminosity
of an \epem\ linear collider, combined with the rise in the
{\epem}$\,\rightarrow\,${\qq} cross section with falling $\sqrt{s}$,
may make it feasible to precisely constrain the evolution
of {\alphas} over a wide range of $Q^2$ in a single experiment.
The execution of such a program 
would have an impact on the design 
of the high energy \epem
collider.
 
Thus, {\epem} annihilation at high energy appears to be a
promising avenue towards the measurement of {\amsmz} to
a relative uncertainty of $\pm 1\,\%$. Furthermore, the high
momentum transfer scale associated with the measurement
({\qsq}$\,\simeq s = (500\,\mathrm{GeV})^2$) makes this
approach an important
component of the program to constrain the
possible anomalous running of~{\alphas}.
For this accuracy to be achievable,
next-to-next to leading order (${\cal{O}}(\alpha_S^3)$)
calculations of {\epem} event shape observables
will be required.
 
\section{{\ppr} ({\pp}) Collisions}

\begin{figure}[htbp]
\centering
\centerline{\psfig{file=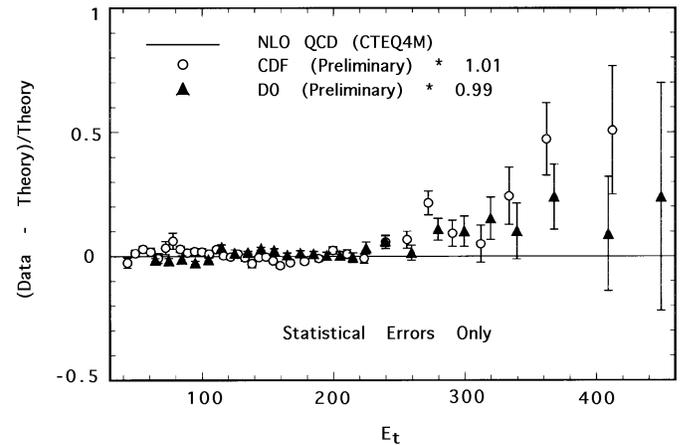,width=8.9cm}}
\caption{The preliminary CDF and D0 Run Ib data compared to
NLO QCD using
CTEQ4M parton distributions. Experimental points normalized
as indicated. This figure is reproduced from Reference~\cite{jetet}.
}
\label{fig:et}
\end{figure}

The greatest potential to extend measurements of {\alphas} to
large values of the momentum transfer scale {\qsq} resides with
the {\ppr} ({\pp}) colliders. 
In adition, many approaches
to the measurement of \alphas in 
{\ppr} ({\pp}) collisions produce a range of values for \alphas\
over a broad lever-arm in $Q^2$. For example, 
the inclusive jet $E_T$
spectrum from the Tevatron extends out to almost 500 GeV
(see Figure~\ref{fig:et}~\cite{jetet}),
providing sensitivity to {\alphas} over a range in momentum transfer
extending from 50 GeV to values nearly
equivalent to that proposed for the next generation of
electron--positron and electron--proton colliders. The LHC,
currently scheduled to begin delivering
beams in the middle of the next decade, will extend this
reach to several~TeV.

On the other hand, measurements of {\alphas} from hadron colliders
have not yet approached the level of accuracy achieved by the most accurate
approaches. For example, a typical approach to constraining
{\alphas} in {\pp} collisions is to study the
ratio of \mbox{$W$ + 1} jet events to \mbox{$W$ + 0}
jet events~\cite{ua2jet}.
Experimental systematics, such as energy scale and resolution
uncertainty, introduce large errors ($\pm \sim 0.015$) in the
value of {\alphas} extracted from this ratio.
In addition, since gluons liberated from the nucleon sea can
themselves form jets, the measurement is sensitive to
the parton distributions used in calculating the $W$ + 1 jet rate.
For the most recent
measurement of this ratio~\cite{d0jet},
the D0 collaboration did not report
a value for {\alphas},
because of an inconsistency between
the best fit value of {\alphas} used in the parton distribution
function and that used in the perturbative matrix element.
 
Current work on the measurement of {\alphas} with
{\pp} data thus concentrates on developing approaches
which remove or reduce the sensitivity of the method to
variations in the parton distribution functions, and to experimental
parameters. This work is still in its early stages, but a number
of promising ideas are being pursued.
 
For example,
a generalization of the $W$ + jet method is the measurement
of the jet cross section ratios
\begin{equation}
R_n^V = {\sigma(V + (n+1) {\rm jets}) \over
\sigma(V + n {\rm jets})} \;\; ,
\end{equation}
where $V$ = $W^{\pm}$,~$Z^0$ is a vector boson.
The UA2 and D0 measurements involved $R_0^V$; for $n \neq 0$,
however, the contribution  from sea gluons, and thus the
dependence on the parton distribution functions, largely
cancels in the ratio. Another approach that is being pursued
is the measurement of the $p_{\perp}$ spectrum of $Z^0$
production -- for this measurement, there is no dependence on
experimental errors such as hadronic energy scales and resolution,
jet algorithm definitions, or hadronization, although the
measurement still requires {\it apriori} knowledge of parton
distribution functions, and only measures {\alphas}
at the single momentum transfer scale
{\qsq}=$M_{Z}^2$.
Finally, fits to the triple differential cross section \cite{d2jet}
\begin{eqalignno}
 { d \sigma \over d E_T d \eta_1 d \eta_2 } &\propto \alpha_S^2
\lbrace  f_{g_1}(x_1) f_{g_2}(x_2) A_{gg}(\eta^{\ast}) 
                 \nonumber  \\[3pt]
&\hskip-1.5cm + f_{g_1}(x_1) {f}_2(x_2) A_{gq}(\eta^{\ast}) +
     {f}_2(x_1) {f}_2(x_2)
     A_{qq}(\eta^{\ast}) \rbrace ;\quad  \\[3pt]
\eta^{\ast} &= { \eta_1 - \eta_2 \over 2} \nonumber\\
x_{1,2} &= {2 E_T\over\sqrt{s}} \left(e^{\pm\eta_1}+e^{\pm\eta_2}\right)
\nonumber
\end{eqalignno}
are being explored, which can simultaneously constrain~{\alphas},
the gluon distribution function $f_g(x)$, and
the non-singlet quark distribution function $f_2(x)$, thus
removing the uncertainty due to poorly constrained parton
distribution functions.
 
All of these studies have only recently been started~\cite{bib-Walter}, 
inspired by the large data samples available with the
completion of Tevatron Run~I. Thus, it will be several
years before the potential for the measurement of
{\alphas} at hadron colliders is fully understood. Finally, it should
be noted
that this method (like any other) requires that at
least
NNLO perturbative calculations
be completed for a determination of {\amsmz} with
$1\,\%$ accuracy. However, most of the matrix elements needed here are
identical to, or are limiting cases of, 
matrix elements necessary for NNLO
calculations of hadronic observables in {\epem} annihilation.
 
\section{Conclusions}
 
Table~\ref{tab:sum} lists the methods we consider promising for
accurate {\alphas} determinations. The items listed above the double
line are established methods for {\alphas} measurements, and we can
evaluate their potential for $1\,\%$ accuracy with reasonable confidence.
The items listed below the double line are either expected to yield
somewhat less accurate determinations of {\alphas}, or they are less
established methods which need further study to better evaluate their
potential for {\alphas} determinations with $1\,\%$ accuracy.

\begin{table*}
\caption[xxx]{Summary of methods for potential $1\,\%$ determinations
of {\amsmz}. The methods listed below the double line are either considered
to yield somewhat less accurate determinations ({\ftwo} at high {\xf} at HERA,
Bjorken sum rule), or they have not yet been fully established ({\pp} and
{\ppr} collisions).}
\label{tab:sum}
\centering
\setlength{\tabcolsep}{8.75pt}
\let\mcol=\multicolumn
\begin{tabular}{|l|l|l|c|l|}
  \hline  &&&&\\[-10pt]
\mcol{1}{|c|}{{\em Process}} & \mcol{1}{|c|}{{\em Approach}} & \mcol{1}{|c|}{{\em NNLO Calculation}} &
    {\em Energy Scale} & \mcol{1}{|c|}{{\em Facilities}} \\[2pt]
  \hline
  \hline &&&&\\[-8pt]
DIS & {\qsq} evolution of {\fthree} & Partial & 2--15 GeV &
               TeV33 fixed target \\
      &   & & 2-45 GeV & LHC fixed target\\[2mm]
DIS &  GLS sum rule   & Available & few GeV & TeV33 fixed target \\
      &   & & $\sim$ 10 GeV  & LHC fixed target\\[2mm]
Hadron spectrum & Spin-averaged $\bb$ and $\cc$ splittings &
      lattice QCD & few--10 GeV & none \\[2mm]
{\epem} & Hadronic observables & Partial & 500 GeV & NLC \\[2pt]
  \hline
  \hline &&&&\\[-8pt]
Polarized DIS & Bjorken sum rule & Available & few GeV & SLAC, DESY, HERA \\
 & & & $\sim$ 10 GeV & NLC fixed target \\[2mm]
DIS & {\qsq} evolution of {\ftwo} at high {\xf} & Partial &
        few--100 GeV &  HERA \\
 & & & $\leq$ 500 GeV & LEP$\times$LHC \\[2mm]
{\pp} & Jet properties & Partial & 100--500 GeV & Tevatron \\
{\ppr} &  & & $\leq$ few TeV & LHC \\[2pt]
  \hline
\end{tabular}
\end{table*}
 
Traditionally, DIS at relatively low momentum transfer has
produced precise determinations of {\alphas}.
In particular, measurements of {\alphas} via
the {\qsq} evolution of {\xfthree} and the GLS sum rule are
expected to each achieve experimental accuracies of 2-3$\,\%$ in the 
upcoming run of the NuTeV Experiment, limited primarily by sample 
statistics, the uncertainty in the calorimeter energy scale,
and the understanding of the composition of the incident neutrino
beam. Thus a high flux tagged neutrino beamline
derived from the {\it full energy} Tevatron primary beam, and
eventually one of the LHC primary proton beams, is a strong
candidate for a facility which will produce a $1\,\%$
measurement of {\alphas} at low {\qsq}.
At present, however, such a facility is not part of the future program
of either laboratory.
We also wish to mention the approach of measuring {\alphas} via the
Bjorken sum rule in polarized deep inelastic scattering.
It is a relatively new method, but could yield a result as accurate as
2-3$\,\%$.
Certain systematic limitations,
such as the corrections for higher twist and the extrapolation
of {\gone} into the unmeasured region at low {\xf}, may be
less problematic if polarized
high energy NLC electron beams are available for fixed
target physics. This issue is worthy of further study.
 
Lattice QCD calculations have matured considerably in the last
few years. First principles calculations of the simplest hadronic
systems, like quarkonia spectra, should be possible with current
technology and computational resources. This implies the potential for
very accurate determinations of {\alphas}
at relatively low {\qsq} from experimental measurements
of the hadron spectrum, with systematic uncertainties largely
independent of all other approaches discussed here.
Since the present experimental errors contribute much less than $1\,\%$,
no future experimental facilities are required for a $1\,\%$ determination
of {\amsmz} from the hadron spectrum.
 
{\epem} annihilation experiments at high center-of-mass energies
are promising for accurate determinations of {\alphas} from measurements
of jet rates and other jet variables. Such experiments could be performed
at an NLC collider with $E_{\rm cm} = 500$ GeV or higher.
 
$ep$ scattering experiments at HERA which measure the structure
function {\ftwo} at high {\xf} over a wide range of {\qsq} can
potentially yield determinations of {\amsmz} with about $2\,\%$ accuracy.
It should be noted that a future LEP$\times$LHC facility can potentially
probe momentum transfers of {\qsq}$\,\sim(500\, \GeV)^2$.
 
Experiments at hadron colliders ({\pp} or {\ppr}) have the
highest potential energy reach. Accurate determinations of {\alphas} will
require either concurrent extractions of the parton distribution
functions (PDFs) from the same experiment or prior knowledge of the PDFs
(with error bars) over the range of {\xf} probed by the process under study.
Feasibility studies are underway now for the Tevatron experiments,
and it is expected that the potential for providing
an accurate measurement of {\alphas} in high energy hadron collisions
will be understood within the next few years. The {\qsq} reach of
the LHC is substantially larger
than that of any other accelerator mentioned here. Should it prove
possible to accurately determine {\alphas} in hadronic collisions,
the construction of a higher energy collider would extend this reach even
further.
 
In summary, the goal of measuring {\alphas} to an accuracy of $1\,\%$, with
a number of complementary approaches, and over a wide range of
{\qsq}, seems feasible. The complete program will likely require
a number of new facilities.
At low {\qsq}, to approach a precision of $1\,\%$ in DIS experiments,
it will most likely be necessary to establish a tagged neutrino beam facility
utilizing the full energy Tevatron beam, or eventually one of the LHC proton
beams. The determination of {\alphas} from the hadron spectrum using lattice
QCD is the only method without facility implications.
At high {\qsq}, a measurement of {\alphas} in {\epem} annihilation
fits quite naturally into the physics program for the proposed Next Linear
Collider. The potential for complementary {\alphas} determinations in {\pp}
collisions at the Tevatron (and {\ppr} collisions at the LHC) still needs
further study, as do measurements of {\ftwo} at high {\xf} at
HERA or at a possible LEP$\times$LHC lepton-hadron collider.
 
 
 

\section*{Acknowledgements}

This work has been supported in part by the Sloan Foundation and
U.S. Department of Energy grants
DE-FG03-94ER40837,
DE-AC03-76SF00515,
DE-FG02-91ER40677,
DE-AC02-76ER03069,
DE-FG02-91ER40685,
DE-FG03-92ER40689.

 
\vfill\break
%
 
%

 
%

\end{document}